%% file: rabbasi_ratevar.tex
\documentclass[superscriptaddress,reprint]{revtex4-2}  

\usepackage{xcolor}
\usepackage{amsmath}    
\usepackage{graphicx}   
\usepackage{dcolumn}    
\usepackage{bm}         
\usepackage[caption=false]{subfig}
\usepackage{hyperref}

\newcommand{\xm}{\relax\ifmmode X_{\mathrm{max}} \else
  $X_{\mathrm{max}}$\fi}
\newcommand{\mxm}{\relax\ifmmode \left<X_{\mathrm{max}}\right> \else
  $\left<X_{\mathrm{max}}\right>$\fi}
\newcommand{\sxm}{\relax\ifmmode \sigma(X_{\mathrm{max}}) \else
  $\sigma(X_{\mathrm{max}})$\fi}
\newcommand{\nm}{\relax\ifmmode N_{\mathrm{max}} \else
  $N_{\mathrm{max}}$\fi}

\begin{document}
\title{Observation of Variations in Cosmic Ray Single Count Rates  During Thunderstorms and Implications for Large-Scale Electric Field Changes}

\input{TA-author-20211010-aastex}


\keywords{cosmic rays --- electric field}

\begin{abstract} 

We present the first observation by the Telescope Array Surface Detector (TASD) of the effect of thunderstorms on the development of  cosmic ray single count rate intensity  over a 700 km$^{2}$ area. Observations of variations in the  secondary low-energy cosmic ray counting rate, using the TASD, allow us to study the electric field inside thunderstorms, on a large scale, as it progresses on top of the 700 km$^{2}$ detector, without dealing with the limitation of narrow exposure in time and space using balloons and aircraft detectors. In this work,  variations in the cosmic ray intensity (single count rate) using the TASD, were studied and found to be on average at the $\sim(0.5-1)\%$ and up to 2\% level. These observations were found to be both in excess and in deficit. They were also found to be correlated with lightning in addition to  thunderstorms. These variations lasted for tens of minutes; their footprint on the ground ranged from 6 to 24 km in diameter and moved in the same direction as the thunderstorm. With the use of simple electric field models inside the cloud and between cloud to ground, the observed variations in the cosmic ray  single count rate were recreated using  CORSIKA simulations. Depending on the electric field model used and the direction of the electric field in that model, the electric field magnitude that reproduces the observed low-energy cosmic ray  single count rate variations was found to be  approximately between 0.2-0.4 GV. This in turn allows us to get a reasonable insight on  the electric field and its effect on cosmic ray air showers inside thunderstorms.

\end{abstract}

\pacs{} 
\maketitle 

\section{Introduction}

Understanding lightning initiation is  one of the most important questions in atmospheric physics.  The heart of the problem of understanding lightning initiation is that, 
with decades of electric fields measurements, the observed values of detected electric field are not sufficient to create a leader or a stroke propagating on a kilometer(s) scale ~\cite{stolz07,winn79}. This could mean that either our understanding of how lightning is initiated or electric field measurements in thunderstorms are inaccurate.

Traditionally, balloons and planes are used to make such measurements. However, there are limitations to obtaining such observations.  At first, sending planes, balloons, and launching rockets inside thunderstorms can be quite difficult and dangerous. Moreover, thunderstorms can span up to square kilometers in size, while the electric field measured by airplanes and balloons spans a small region in comparison. To be in the right location at the right time where the electric field and the potential difference are of a high value can be of low probability. Most importantly, the instrument sent inside a thunderstorm might be responsible for discharging the thunderstorm itself before the electric field has the chance to build up. 

When cosmic ray particles interact in the atmosphere, they  produce a shower of secondary particles.  During thunderstorms, these showers of secondary particles would accelerate or decelerate, depending on their charge and magnitude of the electric field they are propagating through. In principle, studying the effect of the electric field on these secondary particles would allow us to measure and model the electric field in their path indirectly.  

The effect of thunderstorms on  extensive air showers is a hot topic that has been reported on by multiple experiments starting with the Baksan group in 1985~\citep{alex87}. They argued that the effect of the observed cosmic ray variations in the hard and soft components of the shower are due to the electric field in the atmosphere.  Several studies and observations have followed EAS-TOP~\citep{eastop}, Mount Norikura~\citep{mount}, GROWTH~\citep{growth33}, Tibet AS~\citep{tibet}, ARGO-YBJ~\citep{Argo}, and SEVAN~\citep{sevan}, reporting on the  cosmic ray secondary showers (electrons, gamma rays, muons, and neutrons) variation in correlation with thunderstorms.  Most recently, a potential difference of greater than 1 GV inside a cloud ( predicted by C.T.R. Wilson 90 years ago~\cite{wilson}) was indirectly measured in a storm by the Grapes-3 Muon Telescope scientists~\cite{grapes19}.  Such potential difference is almost an order of magnitude larger than the previously reported maximum potential in balloon sounding (0.13 GV)~\citep{mar2001,grapes19}.

In this work, we will present the effect of the electric field in thunderstorms on the extensive air showers  as observed by the Telescope Array Surface Detector (TASD) single count rate.  We will report on 
the observations in  the variation of secondary cosmic-ray single count rate (See the trigger level discussion in Section~\ref{TA}) . The variations are slow, several  kilometers square in area, and moves together with the thunderstorm on top of the 700 km$^2$ detector.  In comparison to detectors that are spread over less than km$^2$ in area (i.e.~\citep{growth33}), 
it is  unclear if the gamma ray emission ceases when the thunderstorm disappears, or when the gamma ray source moves away from the detectors observing the rate variation,  as the thunderclouds moves. We will attempt,  to report on  this question, for the first time, using a large area coverage of 700 km$^2$. Moreover, we will attempt to  interpret this variation, by simulating the effect of the electric field in thunderstorms using multiple simple models. The corresponding increase and decrease of the rate variation in correlation with these models is reproduced and discussed.

\section{The Telescope Array Detector}
\label{TA}
The Telescope Array (TA) detector is located in the southwestern desert of the State of Utah about 1400 m above sea level. Currently it is the largest Ultra High Energy Cosmic Ray (UHECR) experiment in the Northern Hemisphere. The TA detector is comprised of  Surface Detectors (SDs) surrounded by three Fluorescence Detectors (FDs). The main goal of the TA detector is to explore the origin of UHECRs using their energy, composition, and arrival direction.   The FD, which operates on clear moonless nights (approximately 10\% duty cycle) provides a measurement of the longitudinal profile of the Extensive Air Shower (EAS) induced by the primary UHECR, as well as a calorimetric estimate of the EAS energy. The SD part of the detector, with approximately 100\% duty cycle, provides shower footprint information including core location, lateral density profile, and timing, which are used to reconstruct shower geometry and energy.

\begin{figure}
\includegraphics[width=0.25\textwidth]{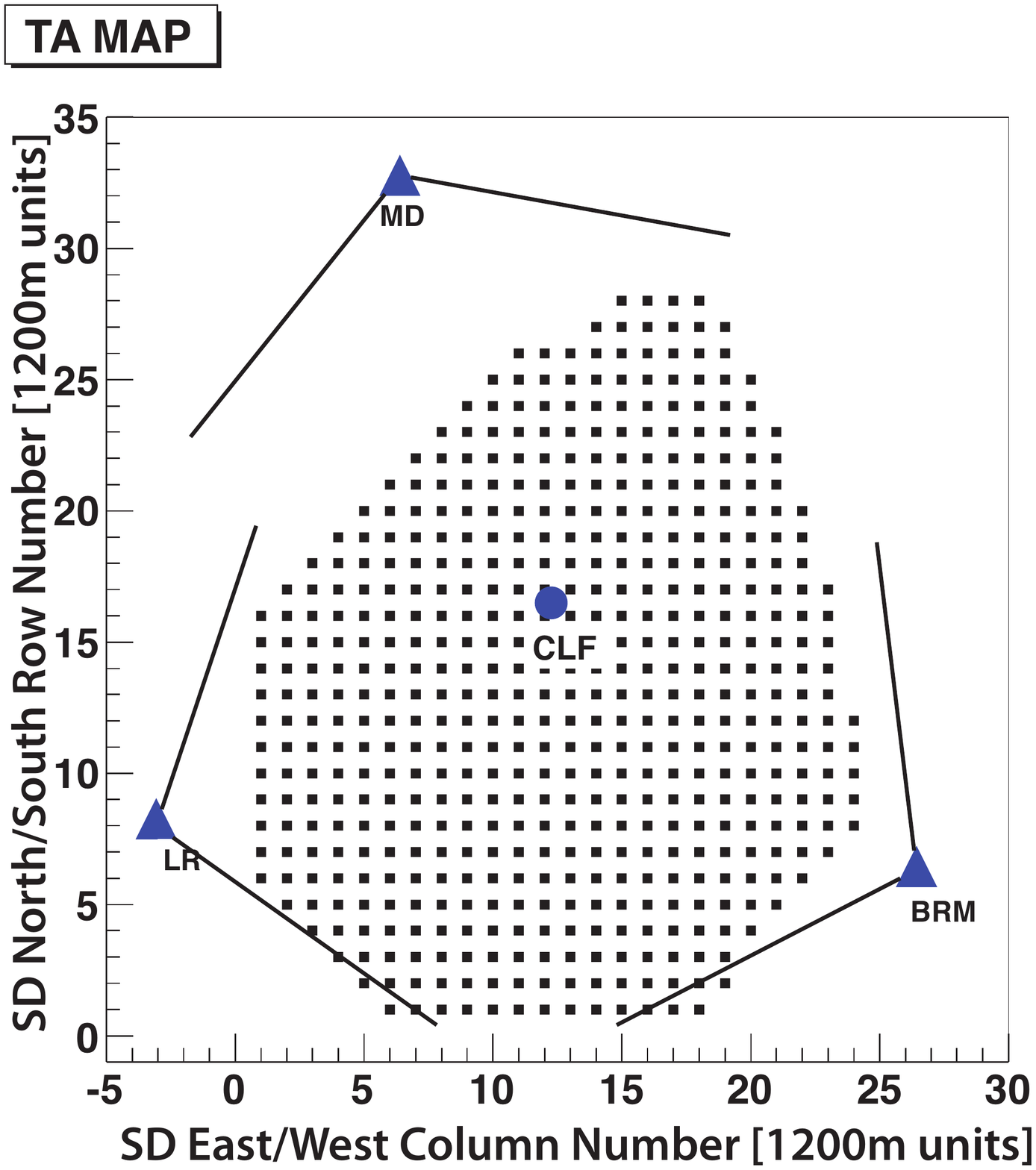}
\hspace{0.5cm}
\includegraphics[width=0.3\textwidth]{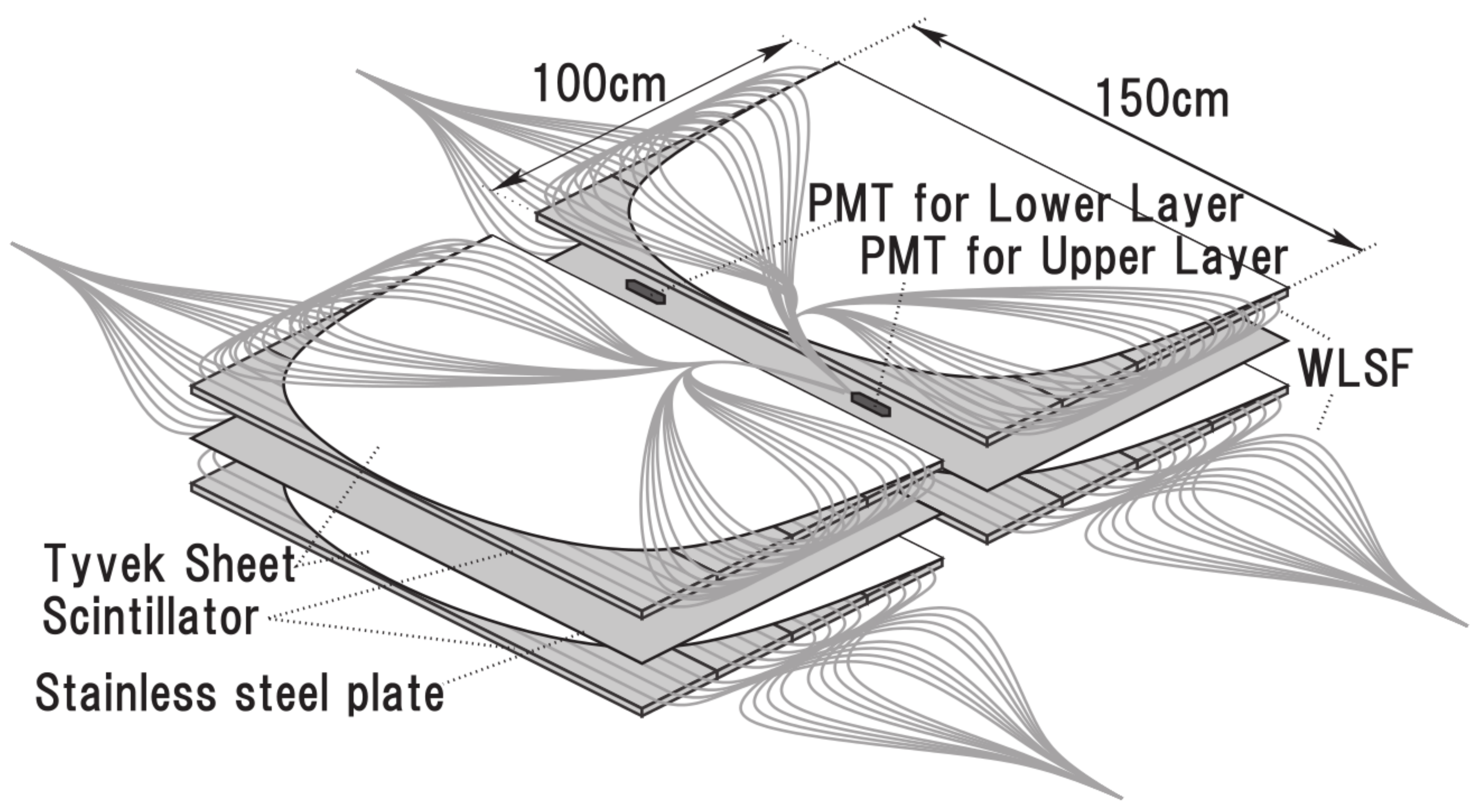}
\caption{{\em top:} The Telescope Array, consisting of 507
scintillator Surface Detectors (SDs) on a 1.2~km grid over a 700~km$^2$ area.  The SD scintillators are enclosed by three fluorescence detectors shown in filled triangles together with their field of view in solid lines. The northernmost fluorescence detector is called Middle Drum while the southern fluorescence detectors are referred to as Black Rock Mesa and Long Ridge. The filled circle in the middle equally spaced from the three fluorescence detectors is the Central Laser Facility used for atmospheric monitoring and detector calibration.
  {\em Bottom:}
 Schematic sketch of the upper and lower 1.2 cm thick plastic scintillator
layers inside the scintillator box, the 1 mm stainless steel plate, the 104 wavelength-shifting (WLS) fibers and
the photomultiplier tubes (PMTs). These items are enclosed in a stainless steel box, 1.5 mm thick on top and
1.2 mm thick on the bottom.~\citep{nonaka2012}. } 
\label{fig:TAmap} 
\end{figure}

The  Surface Detector utilizes plastic scintillators to observe the EAS footprint produced by primary cosmic ray interactions in the atmosphere.  Plastic scintillators are sensitive to all charged particles. The Surface Detector array (SD) part of the TA experiment, is composed of 507 scintillator detectors on a 1.2 km square grid covering  700 km$^{2}$ in area shown in Figure~\ref{fig:TAmap}. Each surface detector houses two layers of plastic scintillator. Each layer of scintillator has an area of 3 m$^2$ and a thickness of 1.2 cm. Each plastic scintillator slab has grooves that has 104 WaveLength-Shifting (WLS) fibers running through them collecting light into PMTs they are bundled and connected to. These scintillator layers are  separated by a 1 mm stainless-steal plate. The  scintillator layers and stainless-steal plate are housed  in light tight, 1.5 mm thick box made of grounded stainless steel (top cover is 1.5 mm thick, with a 1.2 mm thick bottom) under an additional 1.2 mm iron roof providing protection from extreme temperature variations~\citep{nonaka2012}.

  There are a total of three trigger data levels. Level-0, Level-1, and Level-2. Charged particles triggering a single counter (both the upper and the lower scintillators) with an energy above approximately 0.3 Minimum Ionizing Particle (MIP) ($\sim$ 0.75 MeV) are stored in a memory buffer on CPU board as Level-0 trigger data (trigger rate is approximately 750 Hz). 
Charged particles triggering the detector with an energy above approximately 3 MIPs are stored as a level-1 trigger event (trigger rate is approximately 30 Hz).When three adjacent detectors trigger with an energy above 3 MIPs within 8 $\mu$seconds the data is saved as Level-2 trigger (trigger rate is approximately 0.01 Hz).  Level-2 trigger is the one used to study UHECRs and Level-0's main goal is to monitor the health of the detector. In this work we are using  the rate of the detected particles every 10 minutes  recorded by Level-0 trigger dominated by the single particles with primary energy ranging between $\sim 2 \times 10^{10}-10^{13} $ eV.

The TASD is designed to detect the charged components (primarily electrons, positrons, and muons) of the Extensive Air Shower (EAS). The response of the detector has been discussed in detail in~\citep{abbasi2017,nonaka2012}.   Mostly muons and electrons are detected  above approximately 30 MeV. Below this, the total energy deposited by muons and electrons falls off rapidly; below 1 MeV there is no detectable energy deposit as the  electrons fail to penetrate a significant depth into the scintillator~\citep{abbasi2017}.

\section{Observations}

The Telescope Array detector has been in operation since 2008. Thunderstorms continuously pass on top of the Telescope Array detector. In this work, we searched for possible variation in the cosmic ray single count rate using Level-0 trigger in correlation with National Lightning Detection Network (NLDN) activity. There are typically about 750 NLDN 
recorded flashes (intra-cloud and cloud-to-ground) per year over the 700 km$^2$ TASD array. Due to the large number of flashes only days with thunderstorms including a high recorded peak currents  ($>$90 kA) are incorporated in the current search. For the level-0 trigger data collected between 2008-2011, several thunderstorms were observed to produce a variation in the cosmic ray single count rate, the variations were observed during lightning events and  in correlation with thunderstorms in the absence of lightning.

\begin{figure*}[]
\begin{center}
\includegraphics[width=0.7\textwidth]{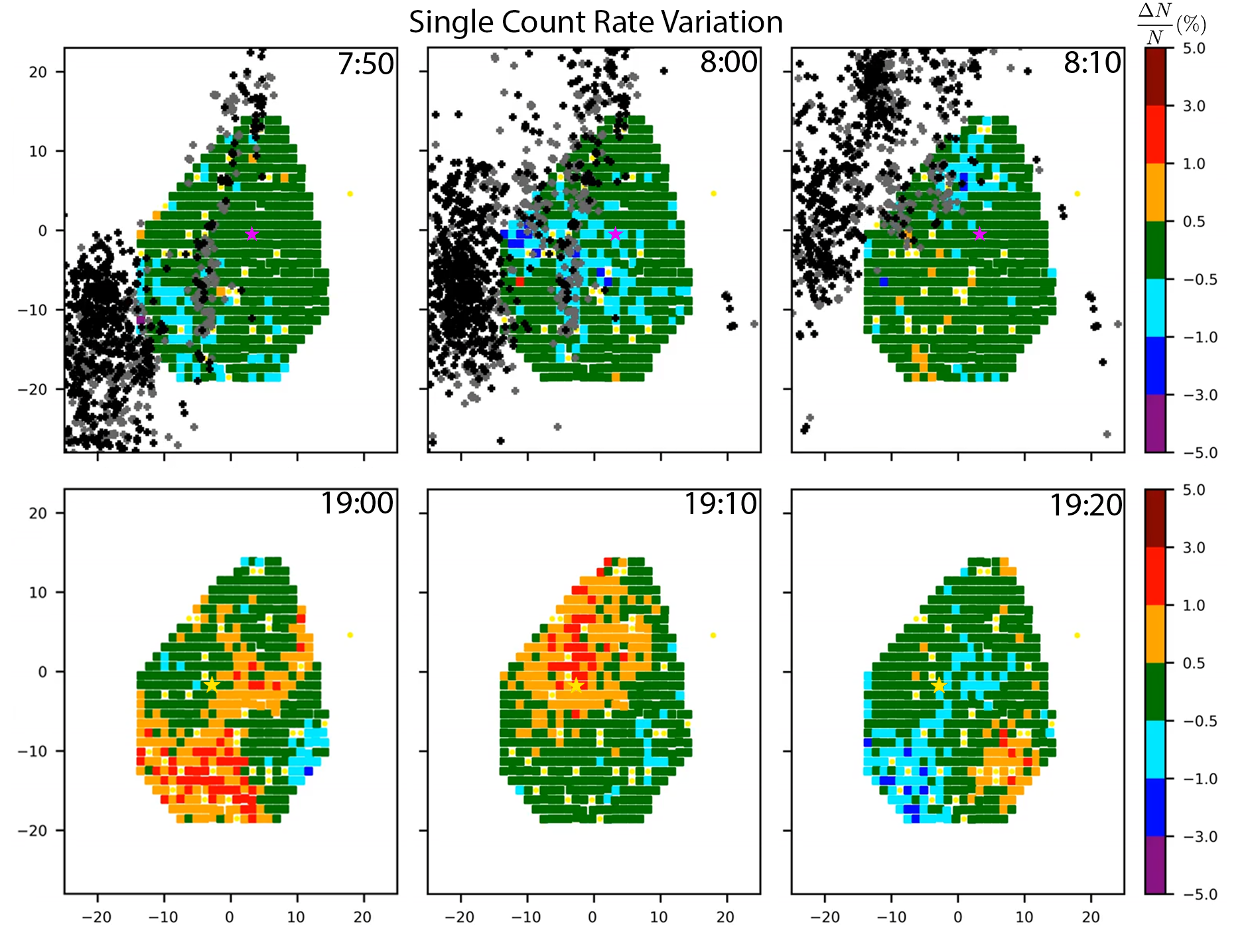}
\caption{Time evolution of the intensity variation of the single count rate change ($\frac{N_c - N_p}{N_p}$)\% or ($\Delta N/ N$)\% on the 09/27/2014 thunderstorm. Each time frame is ten minutes in duration. The starting time in UTC is denoted on each frame. The black and grey crosses marks are the Intra-Cloud and Cloud-to-Ground  lightning sources detected by the NLDN for each frame. The two yellow and pink stars point at the two detectors (1516 (denoted in pink) and 1015 (denoted in yellow)) plotted in Figure~\ref{fig:1d}.  }
\label{fig:obs}
\end{center}
\end{figure*}

As an example, we chose an event observed  on September 27 2014 shown in Figure~\ref{fig:obs}. In Figure~\ref{fig:obs}, each frame lasts for ten minutes in duration.  The time of the start of each frame is denoted on each frame in UTC.  The color scale represents the change of the rate in Level-0 trigger of the current frame $N_c$ from the ten minute frame right before it $N_p$ divided by $N_p$  ($\frac{N_c - N_p}{N_p}$) or ($\Delta N/ N$). Lightning events reported by the NLDN locations are also added in each of the frames in  Figure~\ref{fig:obs} and in the supporting information (SI1). Intra-Cloud in black and Cloud-to-Ground in grey. It is worth noting  that three Terrestrial Gamma-ray  Flashes (TGFs) were reported in~\citep{abbasi2017}  on this day. One of these TGFs was reported at 07:54:35 (during the first frame in Figure~\ref{fig:obs}). 

\begin{figure}[]
\begin{center}
\includegraphics[width=0.45\textwidth]{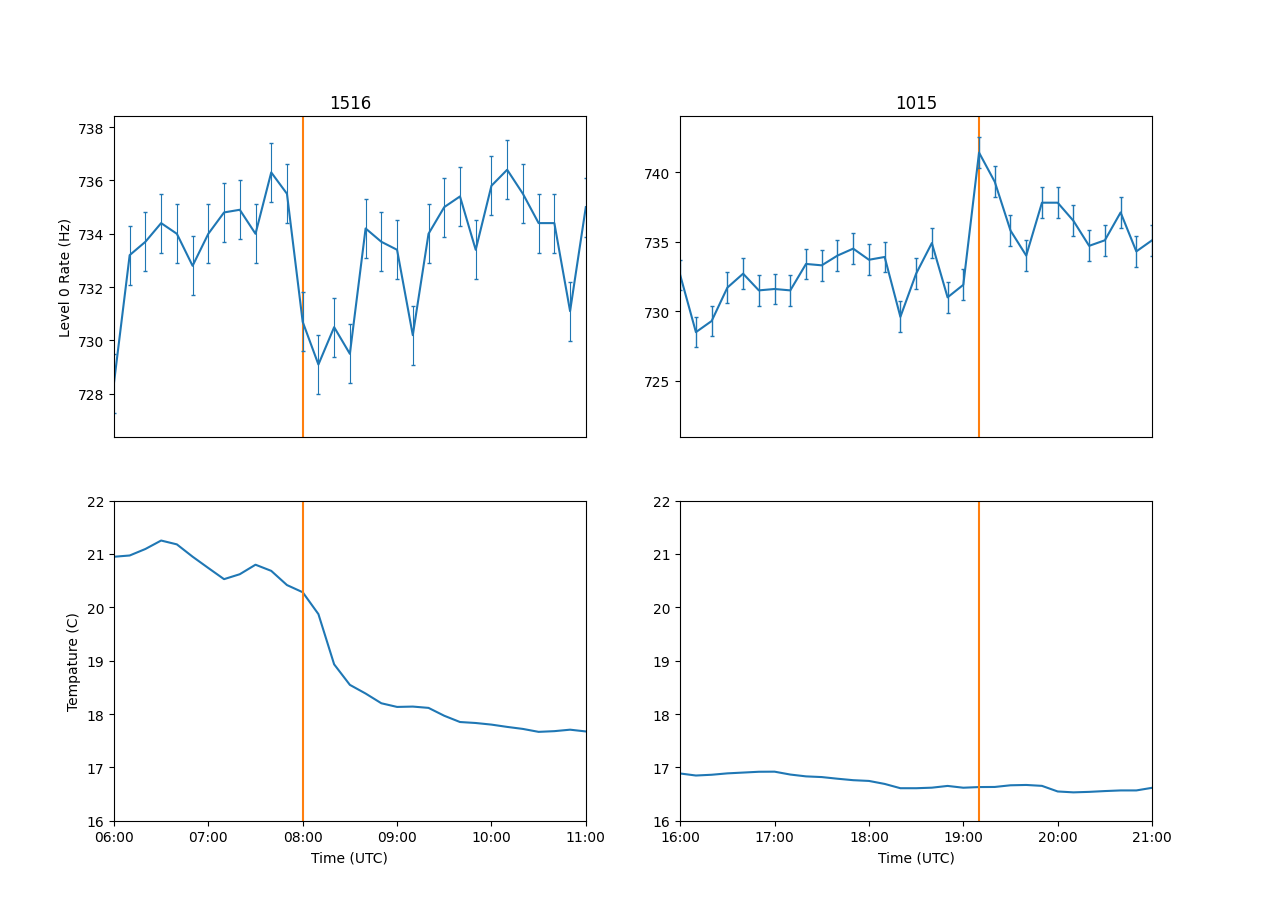}
\caption{Rate variation vs. time and temperature variation vs. time for two detectors numbered (1516 and 1015). Here 1516 shows a deficit in the rate variation (-0.8\%) and 1015 shows an excess in the rate variation (+1.3\%).   }
\label{fig:1d}
\end{center}
\end{figure}

One can see a  movement of a deficit in the intensity variation $\Delta N/ N$ for 30 minutes (from 7:50-8:20 in UTC)  in correlation with lightning activity. In addition, an excess was also found for 30 minutes (from 19:00-19:30 in UTC) in the intensity variation $\Delta N/ N$  during which no lightning activity was reported by NLDN (supporting information, (SI1)). These variations are both seen in correlation with lightning (using NLDN) and thunderstorms (using radar images) in addition or in the absence of lightning (see supporting information videos (SI4, SI5)).  
The variations correlation with pressure is not available at the current time resolution at the ground level. However, the variations were found to be not correlated with temperature changes at the ground level as shown in Figure~\ref{fig:1d} and in the supporting information in (SI2). The size of the variation ranged for this thunderstorm from  6 to 24 km  in diameter on the ground.  The variations were observed in excess and deficit modes  over 10 minutes in durations mostly  between $\pm$(0.5-1)\% and can reach up to 2\% in magnitude. 

\section{CORSIKA Simulations}

The main goal of this simulation work is to  quantify the electric field inside thunderstorms resulting in the observed variations in the single count rate by the TASD detector. To do this we need to learn the conversion of the observed ($\Delta N/N )$ into the equivalent  potential model.  This is done by inserting the atmospheric electric field model into the CORSIKA simulations. Here the CORSIKA package  used in this simulation work is  7.6900~\citep{corsika}, where cosmic rays and their extensive air shower particles propagate through the atmosphere and through the implemented electric field model. Both the electromagnetic  and the muonic components of the showers are traced through the atmosphere and the implemented electric field model until they reach the detector observational level ($\sim$ 1400~m). 

 As a start, two electric field models are used. Note that both models chosen are the simplest electric field models that allow us to reproduce the main observed ($\Delta N/N )$ values. Both models use a uniform  electric field layer. The first model uses a uniform electric field 2 km inside the thundercloud that is located 2 km above ground level. The second model uses a uniform electric field between the thundercloud base and the ground. Both models are illustrated in the Supporting Information (SI3) in the supporting information. In this second model the thundercloud base is 2 km in height from the detector. While thunderstorms structures are known to be complex, both the thundercloud length and height from the ground  used in this work are reasonably representative of thunderstorms at the Southwestern desert of Utah~\citep{abbasi2017}. 

Primary cosmic ray particles composed of protons were generated between 20 GeV -10 TeV.  SIBYLL2.3c~\citep{Sibyl}  is used for the high energy interaction ( $>$ 80 GeV). While,  GHEISHA~\citep{gheisha},  URQMD~\citep{urqmd}, and FLUKA~\citep{Fluka} are used for the low energy model ( $<$ 80 GeV). The zenith and azimuth range from $0^{\circ} \leq \theta \leq 60^{\circ}$ and  $0^{\circ} \leq \phi \leq 360^{\circ}$. The energy threshold of secondary particles were traced until they reach  the following energies: 0.05 GeV for hadrons, 0.5 GeV for muons, 0.001 GeV for electrons and 0.001 GeV for gammas. 

\begin{figure}[]
\begin{center}
 \includegraphics[width=0.49\textwidth]{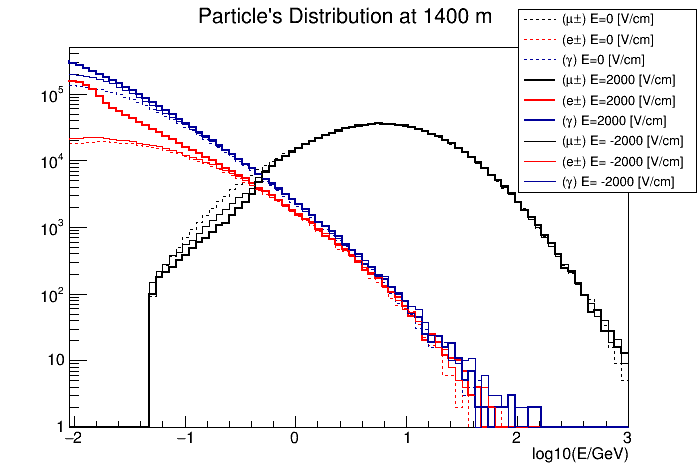}
\caption{The energy distributions of  the muons and electromagnetic components  of the EAS  at 1400~m. The distribution of particles ($e^{\pm}, \mu^{\pm}, \gamma $) included in this plot are without electric field shown in dashed lines for the Cloud-to-Ground model and with electric field of $+$ 2000 V/cm (200 kV/m or 0.4GV/2 km) effect on ($e^{\pm}, \mu^{\pm}, \gamma $) shown in thick solid lines and $-$ 2000 V/cm effect on ($e^{\pm}, \mu^{\pm}, \gamma $) shown in thin solid lines.  Detector response is not included in this distribution. }
\label{fig:Edis}
\end{center}
\end{figure}

\begin{figure*}
\includegraphics[width=0.35\textwidth]{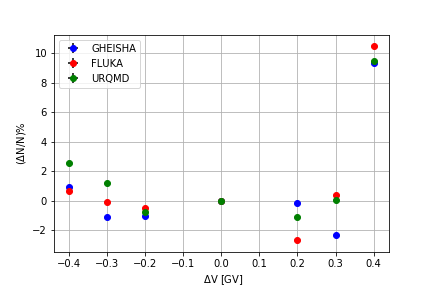}
\includegraphics[width=0.35\textwidth]{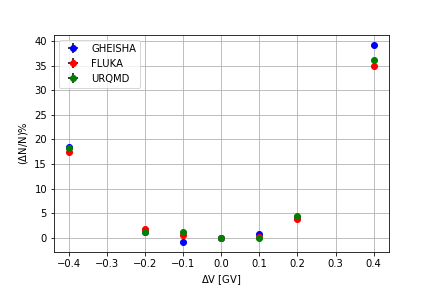}

\caption{{  left: ($ \Delta N/N$)\% vs. $\Delta V$ ,  including statistical error, for  a uniform electric field layer inside the cloud (\textbf{Intra-Cloud model}) using the three low energy model GHEISHA, FLUKA, and URQMD. The  model uses a uniform electric field 2 km  inside the thundercloud that is located 2 km above ground level.  }   { right: ($\Delta N/N$)\% vs $\Delta V$, including statistical error, for a uniform electric field layer between the cloud and ground (\textbf{Cloud-to-Ground model}) using the three low energy model GHEISHA, FLUKA, and URQMD. In this model the thundercloud base is 2 km in height from the detector .}
} 
\label{fig:Epot} 
\end{figure*}

The simulation was  curried out first with no electric field for background. Second, by applying an electric field value that ranges between -2000 to +2000 V/cm (-200 to +200 kV/m). Figure~\ref{fig:Edis} shows the distribution of the electromagnetic ($\gamma$,e$^{\pm}$) and muonic shower components ( $\mu^{\pm}$) on the ground at 1400 m propagated through the atmosphere with electric field at $\pm$ 2000 V/cm and without an electric field from cloud to ground.
The air shower particles  ($\gamma$,e$^{\pm}$, and $\mu^{\pm}$) are then propagated  through the SD detector using an energy dependent response function derived from GEANT4 simulation of the surface detector~\citep{abbasi2017} and following the same trigger condition as the level-0 trigger. The dependence of  ($\Delta N/N)$ on the potential inside the thunderstorms is shown in Figure~\ref{fig:Epot} using both thunderstorm electric field models described in this section.  Note that, the direction of the electric field follows CORSIKA's definition, where positive electric field direction is pointing upwards. 

\section{Discussion}

The simulation results shown in Figure~\ref{fig:Epot} presents ($ \Delta N/N$) vs. the potential difference ($\Delta V$) for both investigated electric field models. The first model included a uniform electric field inside a cloud (\textbf{Intra-Cloud model} ( (SI3) left)) with 2 km in thickness and two kilometers in height from the ground . This model produced both the excess and deficit observed in the variation in the cosmic ray single count rate. While we are unable to distinguish the type of triggering particle from plastic scintillators,  simulations show that the deficit observed by the TASD is dominated by muons. In a negative electric field,  an average deficit using the low energy models (GHEISHA, URQMD, and FLUKA), is  0.75$\pm$0.28\% obtained at $-0.2$ GV. In a positive electric field, an average deficit of $1.3^{+1.17}_{-1.38}$\%  is obtained at $+0.2$ GV. As shown in Figure~\ref{fig:obs} the deficit observed by the TASD is mostly between 0.5 and 1\% and can go up to 2\%.  This observed deficit  is reproduced around $\pm$ 0.2 GV, using this model. 

 As the potential difference increases above 0.3 GV so does the variation in the cosmic ray  single count rate turns from deficit to excess. The excess in the variation of $ \Delta N/N$ strongly depends on the polarity of the electric field inside the thunderstorm in addition to the magnitude of the electric field. Simulations show that while the deficit in muons is stronger with larger potential, an excess in the total number of particles observed by the TASD is expected as the variation of the soft components of the cosmic ray air shower dominates the total number of the observed particles.  It also shows that the observed excess can be obtained depending on the low energy model and polarity. The TASD observed excess is  mostly between 0.5 and 1\% and can go up to 2\%. In a negative electric field an average excess of $1.36^{+1.18}_{-0.44}$\%  is obtained at $-0.4$ GV. In the positive electric field, an average excess of $0.5-2\%$ is obtained with a potential between 0.3 and 0.4 GV.   For the most part,  the magnitude of $\Delta V$  needed to obtain the same observed variation is larger in the negative than in the positive electric field. This asymmetry is due to the fact that the number of electrons exceeds the number of positrons in the extensive air showers. This, in addition to the fact that, there are higher numbers of electrons with lower energies than positrons. Thus the effect of positive fields (accelerating electrons) is larger than the negative field (accelerating positrons)~\citep{Argo}.

The second model included a uniform electric field  of 2 km  in length from the cloud to the ground (\textbf{Cloud-to-Ground model} ((SI3) right)). This model produced only the excess  in the variation in  cosmic ray air single count rate (for the simulation sets produced).  As in the first model, the excess in the total number of particles observed by the TASD is expected as the variation of the soft components of the cosmic ray air shower dominates the total number of observed particles. 
 In a negative electric field, an 
average excess of $1.40^{+0.4}_{-0.2}$\% can be produced by a potential difference of -0.2 GV. In a positive electric field, an excess of 0.5-2\% can be produced by a potential difference of less than 0.2 GV. The excess at a potential difference of -0.4 and 0.4 GV is 20 and 40\% consecutively (much larger than the maximum observed excess of 2\%). Therefore, we conclude that any observed excess resulting from this model is reproduced close to  $\pm$ 0.2 GV in potential.  

 It is important to note that, the interpretation of both models to the observations in the TASD  single count variations is based on the assumption that the duration of the electric field inside the thunderstorm matches that of the duration of the ten minutes recorded observations by the Level-0 filter. However, the duration of the electric field could, in principle, be shorter than 10 minutes and therefore we can assume that our current electric field interpretation is a lower limit value to the possible electric field magnitude that is responsible for the single rate observed variations.

\section{Conclusion}

Variation in the flux of  secondary low-energy cosmic-ray counting rate in association with thunderstorms is reported in this work by the Telescope Array Surface Detector (TASD). The  surface detector utilizes plastic scintillators to observe the charged components (primarily electrons, positrons, and muons) of the cosmic ray air shower.   The variation in  secondary low-energy cosmic-ray counting rate magnitude mostly ranges between (0.5\% and 1\%) and can reach up to 2\%, both in excess and deficit, with a size that range from 6-24 km in diameter. This is the first observation of the variation in the secondary cosmic ray air showers covering 700 km$^2$ in size. Due to the large size of the TASD detector, we can clearly state that the intensity variations in the single count rates observed move in the same direction as the thunderstorms  for tens of minutes at a speed of $\sim 20$ km/10 minutes.  These variations are both seen in correlation with lightning (using NLDN) and thunderstorms (using radar images) in the absence of lightning. 
 
 To interpret the effect of the electric field inside thunderstorms  on the variation  of the cosmic ray secondary shower flux, Monte Carlo simulations are performed with CORISKA. First, cosmic rays air showers are  propagated in multiple electric field models, then  the  secondary shower particles  (both soft and hard components of the shower) are propagated through the detector following the same trigger condition of the data used in this analysis. The total number of particles is then recorded and compared to simulation sets with no electric field.  This simplified models used reproduced both the excess and deficit observed in the variation in the cosmic ray air shower flux. 
 The electric field magnitude found to reproduce  the observed intensity variations was approximately between 0.2-0.4 GV, 
 depending on the electric field model used and the direction of the electric field.  Compared to previous observations, the potential difference recorded by TASD is larger than the  reported maximum potential in balloon sounding (0.13 GV)~\citep{mar2001}.  However, the largest potential difference observed by a cosmic ray detector, thus far, was reported by the Grapes-3 Muon Telescope, with a potential difference of 1 GV~\cite{grapes19}.

In order to interpret the observations of  $ \Delta N/N$  by the TASD, more precisely, it is clear that we need to know the polarity of the thunderstorm. This could in principle be achieved by implementing an array of Electric Field Mills (EFMs) at the Telescope Array site. This  will allow us to better understand the polarity of the observed thunderstorms and therefore model them. Currently, an Electric Field Mill remote station has been installed approximately in the middle of the Telescope Array site for testing. This will enable us to study the relation between SD observations and the development of thunderstorm's  electric field 
 as it progresses on top of the Telescope Array detector.   
\section{Acknowledgements}

 Operation and analyses of this study have been supported by NSF grants AGS-1844306 and AGS-2112709. The Telescope Array experiment is supported by the Japan Society for
the Promotion of Science(JSPS) through
Grants-in-Aid
for Priority Area
431,
for Specially Promoted Research
JP21000002,
for Scientific  Research (S)
JP19104006,
for Specially Promoted Research
JP15H05693,
for Scientific  Research (S)
JP15H05741, for Science Research (A) JP18H03705,
for Young Scientists (A)
JPH26707011,
and for Fostering Joint International Research (B)
JP19KK0074,
by the joint research program of the Institute for Cosmic Ray Research (ICRR), The University of Tokyo;
by the U.S. National Science
Foundation awards PHY-0601915,
PHY-1404495, PHY-1404502, and PHY-1607727;
by the National Research Foundation of Korea
(2016R1A2B4014967, 2016R1A5A1013277, 2017K1A4A3015188, 2017R1A2A1A05071429) ;
by the Russian Academy of
Sciences, RFBR grant 20-02-00625a (INR), IISN project No. 4.4502.13, and Belgian Science Policy under IUAP VII/37 (ULB). The foundations of Dr. Ezekiel R. and Edna Wattis Dumke, Willard L. Eccles, and George S. and Dolores Dor\'e Eccles all helped with generous donations. The State of Utah supported the project through its Economic Development Board, and the University of Utah through the Office of the Vice President for Research. The experimental site became available through the cooperation of the Utah School and Institutional Trust Lands Administration (SITLA), U.S. Bureau of Land Management (BLM), and the U.S. Air Force. We appreciate the assistance of the State of Utah and Fillmore offices of the BLM in crafting the Plan of Development for the site.  Patrick Shea assisted the collaboration with valuable advice  on a variety of topics. The people and the officials of Millard County, Utah have been a source of steadfast and warm support for our work which we greatly appreciate. We are indebted to the Millard County Road Department for their efforts to maintain and clear the roads which get us to our sites. We gratefully acknowledge the contribution from the technical staffs of our home institutions. An allocation of computer time from the Center for High Performance Computing at the University of Utah is gratefully acknowledged. We thank Ryan Said and W. A. Brooks of Vaisala Inc. for providing quality NLDN data lightning discharges over and around the TASD under their academic research use policy.


\clearpage
\section{Supplemental Material}

\begin{figure*}[]
\begin{center}
\includegraphics[width=0.5\textwidth]{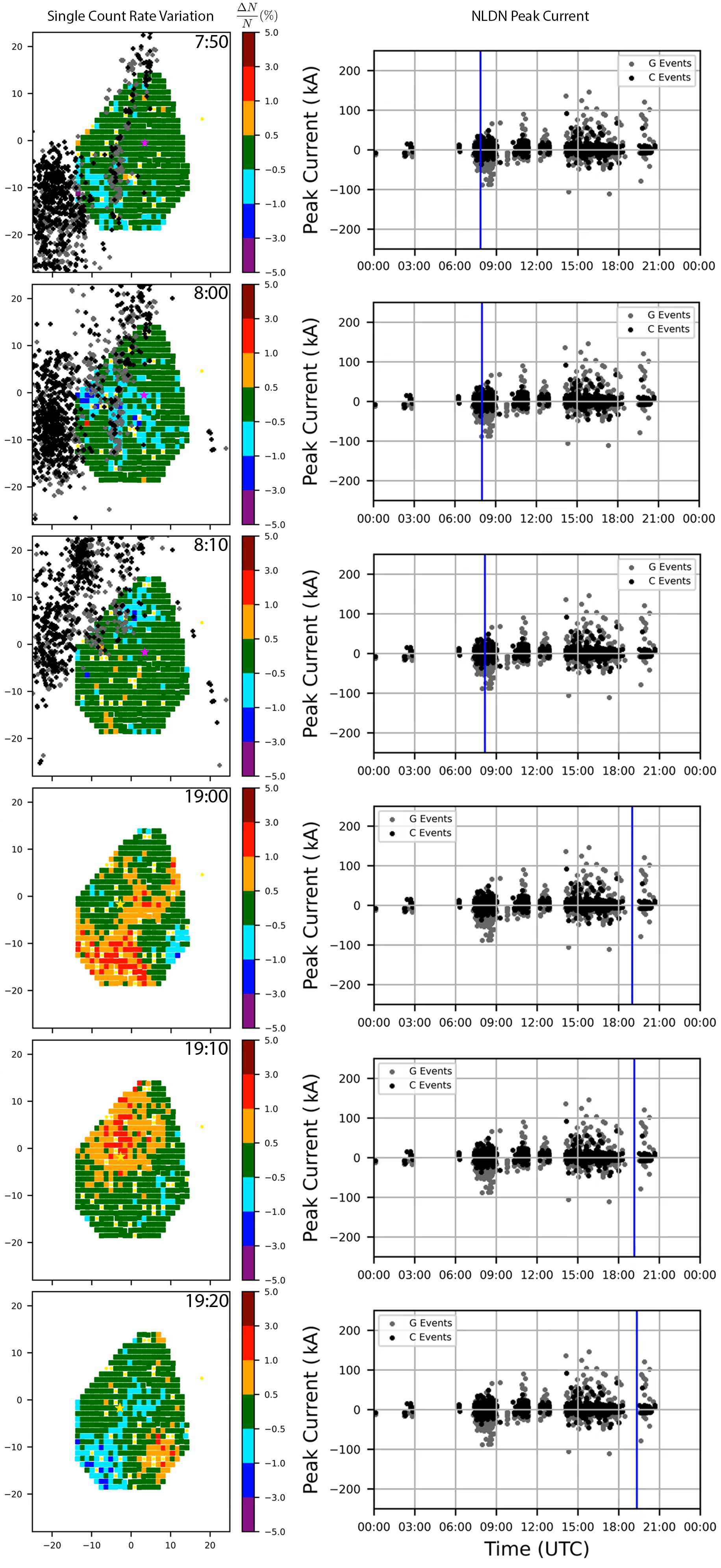}
\caption{\textbf{Supporting Information 1 (SI1):}
left: Time evolution of the intensity variation of the  secondary low-energy cosmic-ray counting rate change ($\frac{N_c - N_p}{N_p}$) or ($\Delta N/ N$) on the 09/27/2014 thunderstorm shown in Figure~\ref{fig:obs}.  Right: NLDN events peak current (kA) vs. time of the day in UTC. The blue line denotes the starting time for each frame on the left hand side. The black and grey cross marks are the Intra-Cloud and Cloud-to-Ground  lightning sources detected by the NLDN for each frame. }
\label{fig:nldn}
\end{center}
\end{figure*}

\begin{figure*}[]
\begin{center}
\includegraphics[width=0.5\textwidth]{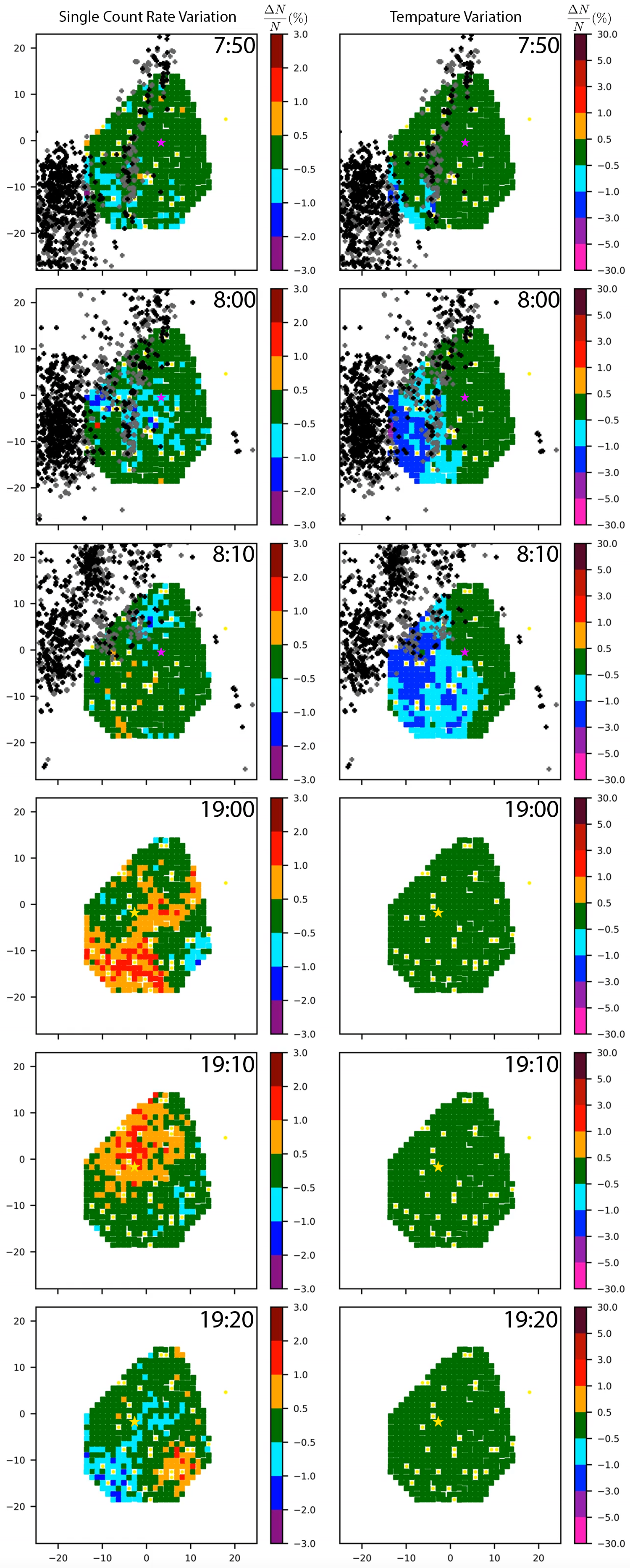}
\caption{\textbf{Supporting Information 2 (SI2):}
left: Time evolution of the intensity variation of the  secondary low-energy cosmic-ray counting rate change ($\frac{N_c - N_p}{N_p}$) or ($\Delta N/ N$) on the 09/27/2014 thunderstorm shown in Figure~\ref{fig:obs}. Right:  Temperature  variation at 1400 m (${T_c - T_p}$) or ($\Delta T$) for the same frames. $T_c$ is the temperature in the current frame and $T_p$ is the temperature in the previous frame.  The starting time is denoted on each frame. The black and grey crosses marks are the Intra-Cloud and Cloud-to-Ground lightning sources detected by the NLDN for each frame. }
\label{fig:temp}
\end{center}
\end{figure*}

\begin{figure*}[]
\begin{center}
\includegraphics[width=0.5\textwidth]{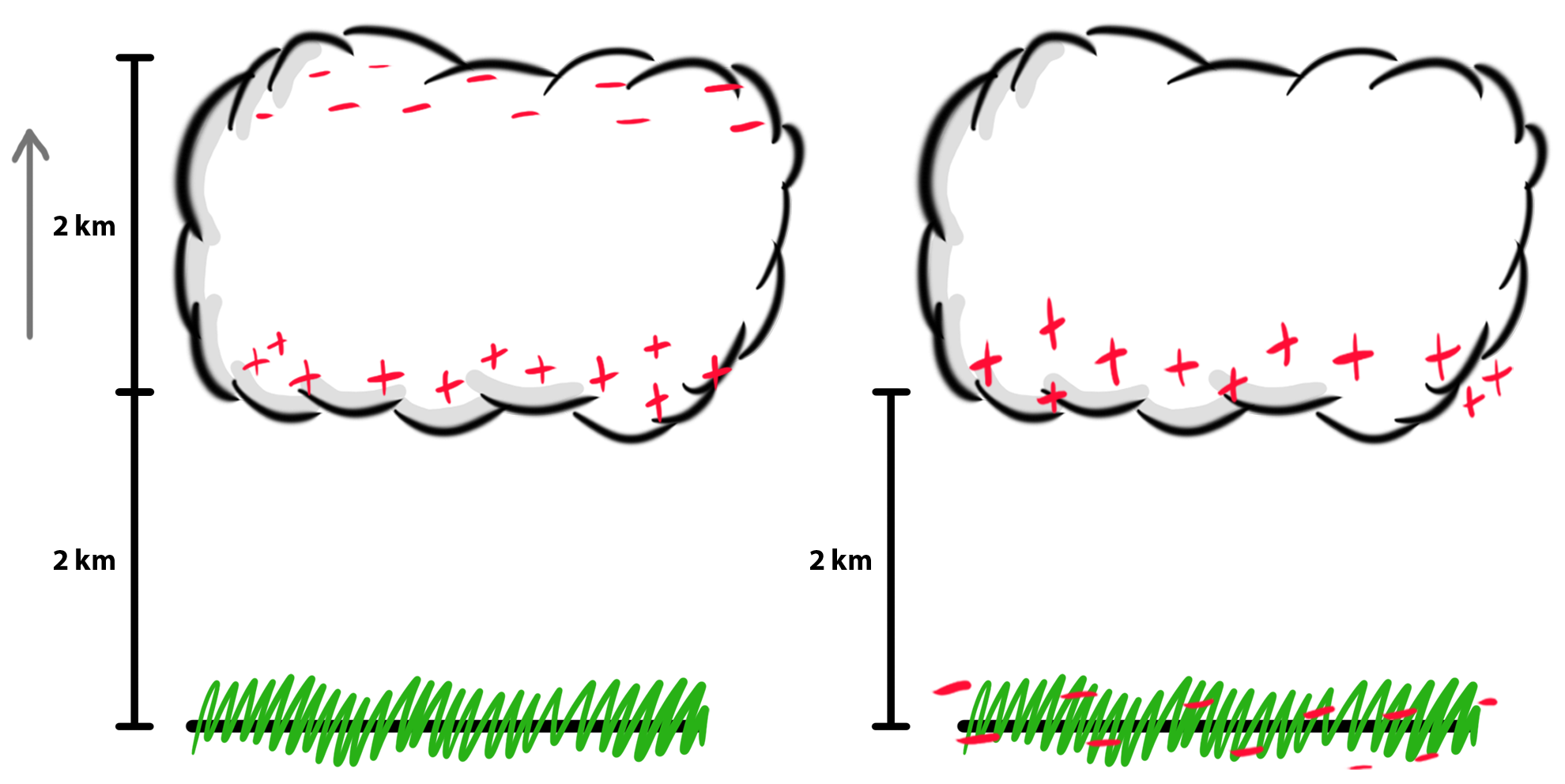}
\caption{\textbf{Supporting Information 3 (SI3):} An illustration of the models used in the simulation  in this work is to quantify the electric field inside thunderstorms resulting in the observed variations in the EAS by the TASD detector. left: The model using a uniform electric field 2 km inside the thundercloud  (\textbf{Intra-Cloud model}) that is located 2 km above ground level. right: The model using a uniform electric field 2 km above ground level (\textbf{Cloud-to-Ground model}). The grey arrow represents the direction of the positive electric field following CORSIKA’s definition, where positive electric field direction is pointing upwards.}
\label{fig:model}
\end{center}
\end{figure*}

\clearpage
\textbf{Supporting Information 4 (SI4):} 
https://youtu.be/608Jm8dujHc. Time evolution of the radar images for the 09/27/2014 thunderstorm from 06:25 - 08:55 including the Telescope Array location marked in red.  The image was extracted from https://www2.mmm.ucar.edu/imagearchive/image

\textbf{Supporting Information 5 (SI5):}
https://youtu.be/V7yIh9wmM30. Time evolution of the radar images for the 09/27/2014 thunderstorm from 18:25 - 19:50 including the Telescope Array location marked in red.  The image was extracted from https://www2.mmm.ucar.edu/imagearchive/image 

\begin{figure*}
\includegraphics[width=1.0\textwidth]{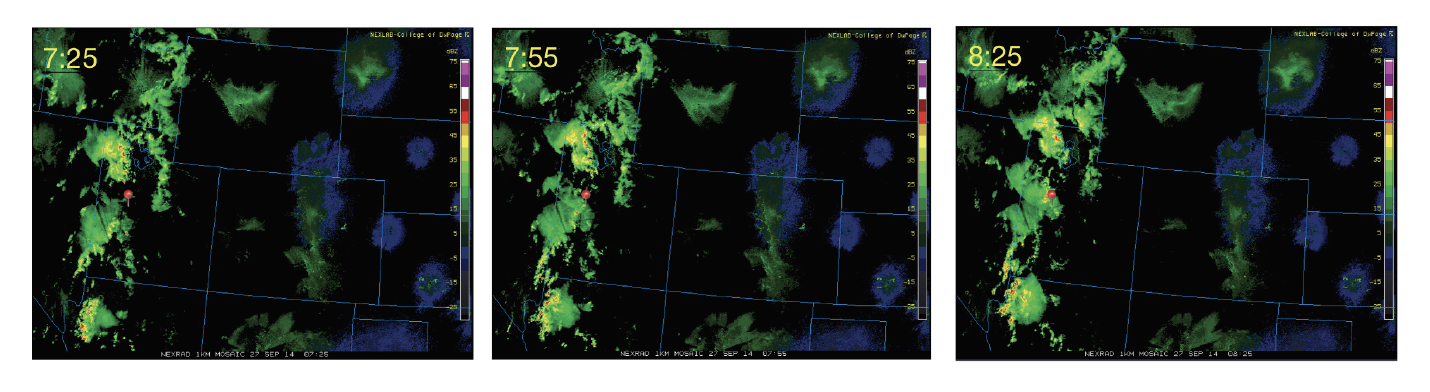}
\includegraphics[width=1.0\textwidth]{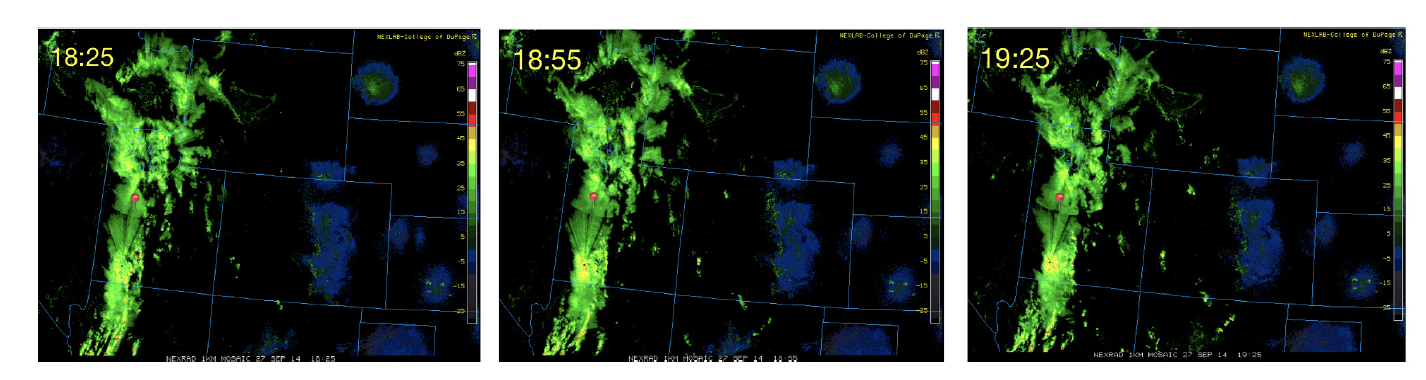}
\caption{ \textbf{Supporting Information 6 (SI6):}Top: Time evolution of the intensity variation of the radar images for the 09/27/2014 thunderstorm from 07:25 - 08:55 including the Telescope Array location marked in red. 
Bottom: Time evolution of the intensity variation of the radar images for the 09/27/2014 thunderstorm from 18:25 - 19:55 including the Telescope Array location marked in red. }
\label{fig:Rad2} 
\end{figure*}

\end{document}

%% file: TA-author-20211010-aastex.tex
\author{R.U. Abbasi}
\email{rabbasi@luc.edu}
\affiliation{Department of Physics, Loyola University Chicago, Chicago, Illinois, USA}

\author{T. Abu-Zayyad}
\affiliation{Department of Physics, Loyola University Chicago, Chicago, Illinois, USA}
\affiliation{High Energy Astrophysics Institute and Department of Physics and Astronomy, University of Utah, Salt Lake City, Utah, USA}

\author{M. Allen}
\affiliation{High Energy Astrophysics Institute and Department of Physics and Astronomy, University of Utah, Salt Lake City, Utah, USA}

\author{Y. Arai}
\affiliation{Graduate School of Science, Osaka City University, Osaka, Osaka, Japan}

\author{R. Arimura}
\affiliation{Graduate School of Science, Osaka City University, Osaka, Osaka, Japan}

\author{E. Barcikowski}
\affiliation{High Energy Astrophysics Institute and Department of Physics and Astronomy, University of Utah, Salt Lake City, Utah, USA}

\author{J.W. Belz}
\affiliation{High Energy Astrophysics Institute and Department of Physics and Astronomy, University of Utah, Salt Lake City, Utah, USA}

\author{D.R. Bergman}
\affiliation{High Energy Astrophysics Institute and Department of Physics and Astronomy, University of Utah, Salt Lake City, Utah, USA}

\author{S.A. Blake}
\affiliation{High Energy Astrophysics Institute and Department of Physics and Astronomy, University of Utah, Salt Lake City, Utah, USA}

\author{I. Buckland}
\affiliation{High Energy Astrophysics Institute and Department of Physics and Astronomy, University of Utah, Salt Lake City, Utah, USA}

% ...
